	\documentclass[
		paper=a4,11pt,american,pagesize,%
		abstract=true,DIV=12,headinclude=true,headlines=0,footinclude=true,footlines=-5,twoside=semi,%
	]{scrartcl}
	\def\docclass{koma}
	\def\version{arxiv}
	\def\draftmode{false} % hide debug info & notes-to-self/todo etc.
\usepackage[utf8]{inputenc}
\makeatletter
\usepackage{xifthen}

\newcommand\iflipics[2]{\ifthenelse{\equal{\docclass}{lipics}}{#1}{#2}}
\newcommand\ifkoma[2]{\ifthenelse{\equal{\docclass}{koma}}{#1}{#2}}
\newcommand\ifieee[2]{\ifthenelse{\equal{\docclass}{ieee}}{#1}{#2}}
\newcommand\ifsiam[2]{\ifthenelse{\equal{\docclass}{siam}}{#1}{#2}}
\newcommand\ifmysiam[2]{\ifthenelse{\equal{\docclass}{my-siam}}{#1}{#2}}
\newcommand\ifacm[2]{\ifthenelse{\equal{\docclass}{acm}}{#1}{#2}}
\newcommand\ifdcc[2]{\ifthenelse{\equal{\docclass}{dcc}}{#1}{#2}}
\ifthenelse{ \equal{\docclass}{lipics} \OR \equal{\docclass}{koma} \OR \equal{\docclass}{ieee} \OR \equal{\docclass}{siam} \OR \equal{\docclass}{my-siam} \OR \equal{\docclass}{acm} \OR \equal{\docclass}{dcc} }{
	\PackageInfo{paper}{Building paper with docclass = \docclass} % all good
}{
	\PackageWarning{paper}{docclass = "\docclass", but must be one of "lipics", "koma", "ieee", "siam", "my-siam", "acm", "dcc"}
}

\newcommand\ifmanuscript[2]{\ifthenelse{\equal{\version}{manuscript}}{#1}{#2}}
\newcommand\ifarxiv[2]{\ifthenelse{\equal{\version}{arxiv}}{#1}{#2}}
\newcommand\ifsubmission[2]{\ifthenelse{\equal{\version}{submission}}{#1}{#2}}
\newcommand\ifproceedings[2]{\ifthenelse{\equal{\version}{proceedings}}{#1}{#2}}
\ifthenelse{ 
	\equal{\version}{manuscript} 
	\OR \equal{\version}{arxiv} 
	\OR \equal{\version}{submission} 
	\OR \equal{\version}{proceedings} 
}{
	\PackageInfo{paper}{Building paper version = \version} % all good
}{
	\PackageWarning{paper}{version = "\version", but must be one of "manuscript", "arxiv", "submission", "proceedings"}
}

\newcommand\ifdraft[2]{\ifthenelse{\equal{\draftmode}{true}}{#1}{#2}}
\ifthenelse{ \equal{\draftmode}{true} \OR \equal{\draftmode}{false} }{
	\PackageInfo{paper}{Building paper with draftmode = \draftmode} % all good
}{
	\PackageWarning{paper}{draftmode = "\draftmode", but must be "true" or "false"}
}

\usepackage[T1]{fontenc}
\ifsiam{
	\usepackage{lmodern}
	\usepackage{slantsc}
}{}
\ifmysiam{
	\usepackage{lmodern}
	\usepackage{slantsc}
}{}
\ifkoma{
	\usepackage{lmodern}
	\usepackage{slantsc}
}{}
\iflipics{
	\usepackage{lmodern}
	\usepackage{slantsc}
}{}
\ifdcc{
	\usepackage{lmodern}
	\usepackage{slantsc}
}{}

\usepackage{babel}
\input{ushyphex.tex} % American English hyphenation exceptions

\usepackage{array,multicol}
\ifieee{
	\usepackage[cmex10]{amsmath,mathtools}
	\usepackage{amsfonts,amssymb}
}{}
\ifkoma{
	\usepackage{amsmath,amsfonts,amssymb,mathtools}
}{}
\iflipics{
	\usepackage{amsmath,amsfonts,amssymb,mathtools}
}{}
\ifsiam{
	\usepackage{amsmath,amsfonts,amssymb,mathtools}
}{}
\ifmysiam{
	\usepackage{amsmath,amsfonts,amssymb,mathtools}
}{}
\ifacm{
	\usepackage{mathtools}
}{}
\ifdcc{
	\usepackage{amsmath,amsfonts,amssymb,mathtools}
}{}

\usepackage{mleftright}\mleftright % redefine left and right to not add space around braces
\usepackage{relsize,xspace,booktabs,adjustbox,needspace,pbox,relsize}
\ifieee{
	
	\usepackage{enumitem}
}{
	\usepackage{enumitem}
}
\usepackage{graphicx}

\ifacm{}{
	\usepackage{colonequals}
}

\usepackage{wref}
\usepackage[bibtex]{url-doi-arxiv}

\newdimen\makeboxdimen
\newcommand\makeboxlike[3][l]{%
\setbox0=\hbox{#2}%
\global\makeboxdimen=\wd0%
\setbox1=\hbox{\makebox[\makeboxdimen][#1]{%
\makebox[0pt][#1]{#3}%
}}%
\ht1=\ht0%
\dp1=\dp0%
\box1%	
}

\newcommand\plaincenter[1]{%
	\mbox{}\hfill#1\hfill\mbox{}%
}

\ifthenelse{\equal{\docclass}{koma} \OR \equal{\docclass}{my-siam}}{
	\setlength\parindent{1.5em}
	\usepackage[headsepline]{scrlayer-scrpage}
	\pagestyle{scrheadings}
	\clearscrheadfoot
	\AtBeginDocument{%
		\automark[section]{}%
	}
	\ohead{\pagemark}
	\rehead{\mytitle}
	\lohead{\headmark}
	\addtokomafont{caption}{\sffamily\small}
	\addtokomafont{captionlabel}{\sffamily\textbf}
	\setcapmargin{2em}
}{}
\ifthenelse{\equal{\docclass}{my-siam}}{
	\setcapmargin{1em}
	\setcapindent{0em}
}{}
\ifmysiam{
	\setlength\parskip{0pt}
	\RedeclareSectionCommand[
		beforeskip=-1.25\baselineskip,
		afterskip=0.75\baselineskip,
	]{section}
	\RedeclareSectionCommand[
		beforeskip=-1\baselineskip,
		afterskip=-1.5em,
	]{subsection}
	\RedeclareSectionCommand[
		beforeskip=-1\baselineskip,
		afterskip=-1.5em,
	]{subsubsection}
	\RedeclareSectionCommand[
		beforeskip=-.25\baselineskip,
		indent=1.5em,
		afterskip=-1em,
	]{paragraph}
}{}
\ifdcc{
	\ifproceedings{}{
		\pagestyle{plain}
		\setlength{\footskip}{5ex}
	}
}{}

\AtBeginDocument{%
	\let\mytitle\@title%
}

\let\oldthebibliography\thebibliography
\renewcommand\thebibliography[1]{%
	\oldthebibliography{#1}%
	\pdfbookmark[1]{References}{}%
}

\usepackage{lscape} % sideways figures

\ifkoma{
	\usepackage{float}
	\floatstyle{plain}
	\usepackage{newfloat}
	\DeclareFloatingEnvironment[%
			name=Algorithm,%
			placement=thb,%
		]{algorithm}
}{}
\iflipics{
	\usepackage{newfloat}
	\DeclareFloatingEnvironment[%
			name=Algorithm,%
			placement=thb,%
		]{algorithm}
}{}
\ifmysiam{

	\usepackage{newfloat}
	\DeclareFloatingEnvironment[%
			name=Algorithm,%
			placement=thb,%

		]{algorithm}
}

\ifdcc{
	\usepackage{float}
	\usepackage[font={small},labelfont=bf]{caption}
}{}
\ifmysiam{
	\setcounter{topnumber}{3}
	\setcounter{bottomnumber}{3}
	\setcounter{totalnumber}{3}     % 2 may work better
	\setcounter{dbltopnumber}{3}    % for 2-column pages
}{}
\ifsiam{
	\setcounter{topnumber}{3}
	\setcounter{bottomnumber}{3}
	\setcounter{totalnumber}{3}     % 2 may work better
	\setcounter{dbltopnumber}{3}    % for 2-column pages
}{}

\usepackage{dcolumn}

\usepackage{textcomp} % needed for upquote
\usepackage{listings}

\usepackage{tikz}

\usetikzlibrary{positioning,arrows.meta,fit}
\usetikzlibrary{backgrounds,calc,trees,graphs}
\usetikzlibrary{shapes.geometric,shapes.misc}

\pgfdeclarelayer{background}
\pgfsetlayers{background,main}

\iflipics{
	\newtheorem{fact}[theorem]{Fact}

	\newenvironment{proofof}[1]{%
		\begin{proof}[{{Proof of #1{}}}]%
	}{%
		\end{proof}%
	}
}{}
\ifacm{
	\AtEndPreamble{
		\theoremstyle{acmdefinition}
		\newtheorem{remark}[theorem]{Remark}
		\newtheorem{fact}[theorem]{Fact}
	}
	
	\newenvironment{proofof}[1]{%
		\begin{proof}[{{Proof of #1{}}}]%
	}{%
		\end{proof}%
	}
}{}
\ifsiam{
	\newtheorem{remark}{Remark}
	\newenvironment{proofof}[1]{%
			\begin{proof}[{{#1{}}}]%
		}{%
			\end{proof}%
		}
}{}
\ifthenelse{\equal{\docclass}{lipics} \OR \equal{\docclass}{siam} \OR \equal{\docclass}{acm}}{}{
	\usepackage[amsmath,hyperref,thmmarks]{ntheorem}
	
	\theorembodyfont{\slshape}
	\theoremseparator{:}
	\newtheoremstyle{proofstyle}%
	  {\item[\theorem@headerfont\hskip\labelsep ##1\theorem@separator]}%
	  {\item[\theorem@headerfont\hskip\labelsep ##3\theorem@separator]}
	
	\theorempreskip{\topsep} % new version of ntheorem

	\theoremsymbol{\adjustbox{scale=.8}{$\triangleleft\mkern-1mu$}}
	
	\theoremstyle{plain}
	\theorempreskip{\topsep}

	\theoremstyle{plain}
	\theorembodyfont{\upshape}

	\theoremsymbol{\raisebox{-.25ex}{$\Box$}}
	\qedsymbol{\raisebox{-.25ex}{$\Box$}}
	
	\theoremstyle{proofstyle}
	\newtheorem{proof}{Proof}

}

\iflipics{
	\newenvironment{thmenumerate}[2][]{%
		\begin{enumerate}[
			label={\textsf{\textbf{\color{darkgray}{\makebox[\widthof{(a)}][c]{\textup{(\alph*)}}}}}},
			ref={\ref{#2}\kern.1em--\kern.1em(\alph*)},
			itemsep=0pt,
			topsep=.5ex,
			leftmargin=1.75em,
			#1
		]%
	}{%
		\end{enumerate}%
	}
}{
	
}

\newcommand*\ie{\mbox{i.\hspace{.2ex}e.}}
\newcommand*\eg{\mbox{e.\hspace{.2ex}g.}}

\usepackage{fixmath}

\newcommand{\ESymbol}{\mathbb{E}}

\newcommand{\ProbSymbol}{\ensuremath{\mathbb{P}}}

\DeclarePairedDelimiterXPP\Prob[1]{\ProbSymbol}[]{}{%
	#1%
}
\DeclarePairedDelimiterXPP\E[1]{\ESymbol}[]{}{%
	#1%
}
\DeclarePairedDelimiterXPP\Eover[2]{\ESymbol_{#1}}[]{}{%
	#2%
}
\DeclarePairedDelimiterXPP\ProbIn[2]{\ProbSymbol_{#1}}[]{}{%
	#2%
}
\providecommand{\Prob}{} % hack for syntax highlighting ...
\providecommand{\ProbIn}{} % hack for syntax highlighting ...
\providecommand{\E}{} % hack for syntax highlighting ...
\providecommand{\Eover}{} % hack for syntax highlighting ...

\newcommand{\surroundedmath}[3]{% #1=mathrel/mathbin/etc #2=spacing #3=symbol
	\mathchoice{%display
		#1{#2{#3}#2}%
	}{%text
		#1{#3}%
	}{%script
		#1{#3}%
	}{%scriptsript
		#1{#3}%
	}%
}
\newcommand\rel[1]{\surroundedmath{\mathrel}{\:}{#1}}
\newcommand\wrel[1]{\surroundedmath{\mathrel}{\;}{#1}}
\newcommand\wwrel[1]{\surroundedmath{\mathrel}{\;\;}{#1}}

\iflipics{}{
	\makeatletter
	\let\oldalign\align
	\let\endoldalign\endalign
	\renewenvironment{align}{%
		\begingroup%
		\let\oldhalign\halign
		\def\halign{%
			\let\oldbreak\\%
			\def\nonnumberbreak{\nonumber\oldbreak*}%
			\def\\{%
				\@ifstar{\nonnumberbreak}{\oldbreak}%
			}%
			\oldhalign%
		}
		\oldalign%
	}{%
		\endoldalign%
		\endgroup%
	}
}
\newcommand*\numberthis[1][]{\stepcounter{equation}\tag{\theequation}}

\allowdisplaybreaks[3]

\newcommand\splitaftercomma[1]{%
  \begingroup
  \begingroup\lccode`~=`, \lowercase{\endgroup
    \edef~{\mathchar\the\mathcode`, \penalty0 \noexpand\hspace{0pt plus .25em}}%
  }\mathcode`,="8000 #1%
  \endgroup
}

\def\mydots{\xleaders\hbox to.5em{\hfill.\hfill}\hfill}
\newlength\tmpLenNotations

\ifdraft{%
    \usepackage[switch]{lineno} 
    \linenumbers
	\overfullrule=6mm
	
	\usepackage[color,notref,notcite]{showkeys}
	\definecolor{refkey}{gray}{.99}
	\colorlet{labelkey}{green!60!black!60}
	
	\usepackage[inline,nolabel]{showlabels}

	\showlabels{cite}
	\showlabels{citealt}
	\showlabels{citealp}
	\showlabels{citet}
	\showlabels{Citet}
	\showlabels{citep}
	\showlabels{citeauthor}
	\showlabels{Citeauthor}
	\showlabels{citefullauthor}
	\showlabels{citeyear}
	\showlabels{citeyearpar}
	\showlabels{wref}
	\showlabels{wpref}
	\showlabels{wtpref}
	\showlabels{wildref}
	\showlabels{wildpageref}
	\showlabels{wildtpageref}
}{}

\iflipics{
	\ifmanuscript{\hideLIPIcs}{}
	\ifarxiv{\hideLIPIcs}{}
	\ifsubmission{}{\nolinenumbers}
}{}

\ifdraft{}{%
	\usepackage{microtype}
}

\hypersetup{
	final,
	unicode=true, 
	bookmarks=true,
	bookmarksnumbered=true,
	bookmarksdepth=2,
	bookmarksopen=true,
	breaklinks=true,
	hidelinks,
}

\newsavebox\tmpbox

\ifdcc{
	\renewcommand\paragraph{\@startsection{paragraph}{4}{\parindent}%\z@}%
	                                      {\smallskipamount}%{3.25ex \@plus1ex \@minus.2ex}%
	                                      {-1em}%
	                                      {\normalfont\normalsize\bfseries}}
}{}
\iflipics{
	\let\oldparagraph\paragraph
	\renewcommand\paragraph[1]{%
		\oldparagraph*{#1}
	}
}{
	\let\oldparagraph\paragraph
	\renewcommand\paragraph[1]{%
		\oldparagraph{#1.}
	}
}

\ifmysiam{
	\let\oldsubsection\subsection
	\renewcommand\subsection[1]{%
		\oldsubsection{#1.}%
	}
	\let\oldsubsubsection\subsubsection
	\renewcommand\subsubsection[1]{%
		\oldsubsubsection{#1.}%
	}
}{}
\ifsiam{
	\let\oldsubsection\subsection
	\renewcommand\subsection[1]{%
		\oldsubsection{#1.}%
	}
	\let\oldsubsubsection\subsubsection
	\renewcommand\subsubsection[1]{%
		\oldsubsubsection{#1.}%
	}
}{}

\let\epsilon\varepsilon

\def\myacknowledgements{}
\ifkoma{
	
}{}
\ifieee{
	
}{}
\ifsiam{
	
}{}
\ifmysiam{
	
}{}
\ifdcc{
	
}{}
\ifacm{
	
}{}

\newcommand\pair[2]{\genfrac{[}{]}{0pt}{1}{#1}{#2}}
\newcommand\terminalPair[2]{\pair{\texttt{#1}}{\texttt{#2}}}

  \newcommand\tao{{\text{\small$\terminalPair{A}{(}$}}}
  \newcommand\tco{{\text{\small$\terminalPair{C}{(}$}}}
  \newcommand\tgo{{\text{\small$\terminalPair{G}{(}$}}}
  \newcommand\tuo{{\text{\small$\terminalPair{U}{(}$}}}
  \newcommand\tac{{\text{\small$\terminalPair{A}{)}$}}}
  \newcommand\tcc{{\text{\small$\terminalPair{C}{)}$}}}
  \newcommand\tgc{{\text{\small$\terminalPair{G}{)}$}}}
  \newcommand\tuc{{\text{\small$\terminalPair{U}{)}$}}}
\renewcommand\tau{{\text{\small$\terminalPair{A}{\textbullet}$}}}
  \newcommand\tcu{{\text{\small$\terminalPair{C}{\textbullet}$}}}
  \newcommand\tgu{{\text{\small$\terminalPair{G}{\textbullet}$}}}
  \newcommand\tuu{{\text{\small$\terminalPair{U}{\textbullet}$}}}

\newcommand\secondaryStructureSymbol[1]{\makeboxlike{\texttt{M}}{\texttt{\bfseries #1}}\xspace}
\newcommand\ssu{\secondaryStructureSymbol{\textbullet}} %% secondary structure unpaired
\newcommand\sso{\secondaryStructureSymbol{(}} %% secondary structure open
\newcommand\ssc{\secondaryStructureSymbol{)}} %% secondary structure close

\newcommand\leftmost[2]{\mathrm{lmd}_{#1}(#2)}

\usepackage{placeins} % for \FloatBarrier

\makeatother

\usepackage{hyperref}
    \title{Towards Optimal Grammars for RNA Structures}
    \author{
        Evarista Onokpasa%
    \thanks{%
        University of Liverpool, UK,
        \texttt{\{evarista.onokpasa,\,sebastian.wild,\,pwong\}\,@\,liverpool.ac.uk}
    }
    \and
        Sebastian Wild$^\ast$
    \and
        Prudence W.H.\ Wong$^\ast$
    }

	\date{\small\today}

\begin{document}

\maketitle

\begin{abstract}
In past work (Onokpasa, Wild, Wong, DCC 2023), we showed that (a) for joint compression of RNA sequence and structure, stochastic context-free grammars are the best known compressors and (b) that grammars which have better compression ability also show better performance in \textit{ab initio} structure prediction. 
Previous grammars were manually curated by human experts.
In this work, we develop a framework for automatic and systematic search algorithms for stochastic grammars with better compression (and prediction) ability for RNA. We perform an exhaustive search of small grammars
and identify grammars that surpass the performance of human-expert grammars.
\end{abstract}

\section{Introduction}

\begin{figure}[tbp]
	\bigskip
	\begin{minipage}[t]{.35\textwidth}
		~\\[-\baselineskip]
		\resizebox{.95\linewidth}!{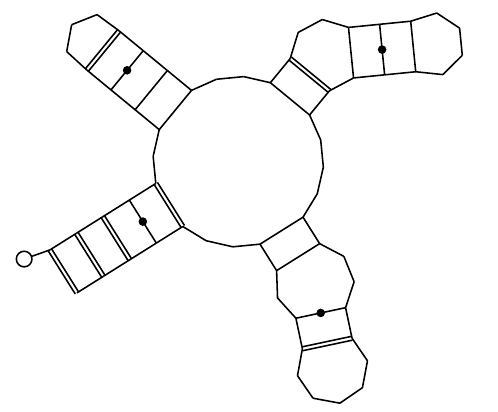}
	\end{minipage}%
	\begin{minipage}[t]{.65\textwidth}
		~\\[-3ex]
		\hspace*{-2em}\resizebox{1.1\linewidth}!{\begin{tikzpicture}[yscale=.2,xscale=.18]
			\foreach \x in {1,...,62} { \node[circle,draw=black,fill=black,inner sep=.8pt] (b\x) at (\x,0) {} ; }
			\foreach \f/\t/\h in {%
				2/62/9,3/61/8.5,4/60/8,5/59/7.5,6/58/7,%
				8/18/3.5,9/17/3,10/16/2.5,11/15/2,%
				21/36/4.5,22/35/4,25/34/3,26/33/2.5,27/32/2,%
				40/55/4,41/54/3.5,44/52/2.5,45/51/2
			} 
			{
				\draw[semithick] (b\f) .. controls ($(b\f)+(0,\h)$) and ($(b\t)+(0,\h)$) .. (b\t) ; 
				\path (b\f) ++(0,-4.2) node[anchor=base] {\scriptsize\sso} ;
				\path (b\t) ++(0,-4.2) node[anchor=base] {\scriptsize\ssc} ;
			}
			\foreach \u in {1,7,12,13,14,19,20,29,23,24,28,29,30,31,37,38,39,42,43,46,47,48,49,50,53,56,57} {
				\path (b\u) ++(0,-4.2) node[anchor=base] {\scriptsize\ssu} ;
			}
			\draw (b1) -- (b62);
			\foreach \i in {1,5,10,...,62,62} {
				\node[scale=.45] at ($(b\i)+(0,-1)$) {\i} ;
			}
			\foreach [count=\i] \b in {G,C,C,C,U,G,A,U,A,G,C,G,U,A,G,U,U,A,C,U,A,G,C,G,A,G,U,C,U,G,U,A,U,U,C,U,A,A,G,A,A,G,A,U,C,A,C,U,G,A,G,G,G,U,U,C,G,C,G,G,G,G} {
				\node[scale=.5] at ($(b\i)+(0,-2.6)$) {\texttt{\b}} ;
			}
			
		\end{tikzpicture}}
		\vspace*{-3ex}
		\caption{%
			An example RNA sequence and structure.
			\textbf{Left:} schematic drawing of structure. 
			\textbf{Above:} Representation as dot-bracket sequence when the backbone is ``pulled straight''.
		}%
		\label{fig:rna-example}
	\end{minipage}
	\vspace*{-2ex}
\end{figure}

In this paper, we study the fundamental question of 
capturing typical folding structures of RNA molecules. 
Ribonucleic acid (RNA) is a bio-polymer that serves various roles in the coding, decoding, expression and regulation of genes in cells. 
An RNA molecule consists of a chain of \emph{nucleotides} each having a \emph{base} attached to it (either adenine (\texttt A), cytosine (\texttt C), guanine (\texttt G), or uracil (\texttt U)); this string of bases forms the \emph{sequence} of the molecule.
Unlike the related DNA, RNA is usually single-stranded and forms spatial structures by folding onto itself (similar to proteins), with complementary bases forming stabilizing hydrogen bonds.
The (well-nested) set of (indices of the) bases that form such pairs is the \emph{secondary structure} of the molecule;
it can be encoded by the dot-bracket notation,
see \wref{fig:rna-example}; a formal definition is given in \wref{sec:preliminaries}.

The secondary structure is instrumental for the biological function of non-coding RNA molecules and of great interest to biologists.
Much research has hence been devoted to computationally \emph{predicting} the secondary structure 
from a known RNA sequence (\emph{ab initio} RNA secondary-structure prediction)%
~\cite{DurbinEddyKroghMitchison1998,GorodkinRuzzo2014,TurnerMathews2016}.
In our recent work~\cite{OnokpasaWildWong2023}, we showed that joint compression of RNA sequence and structure data can serve as a robust proxy for the prediction quality of different stochastic context-free grammar (SCFG) models of RNA secondary structures, a state-of-the-art formalism for \emph{ab initio} structure prediction.
We also showed that the RNA-specific SCFG-based compression outperforms by far the best general-purpose compressors such as \emph{paq8l} (\url{http://mattmahoney.net/dc/\#paq})
on RNA data.

In~\cite{OnokpasaWildWong2023}, we use SCFGs designed resp.\ collected by human domain experts~\cite{DowellEddy2004,LiuYangChenBuZhangYe2008,NebelScheid2011}.
\textbf{In this work, we move} from  these isolated examples \textbf{towards a framework for systematic and automated search for \emph{optimal} grammars.}
By comparing SCFGs using their achieved compressed size instead of structure prediction performance, we eradicate intricacies and ongoing debates of how to measure the distance between predicted and true secondary structures~\cite{Mathews2019}; 
our work thus paves the way for a well-defined open contest on finding SCFGs that best capture the essence of stable (minimum-free-energy) RNA structures.

On the technical side, we provide reference implementations of all key components needed for testing and evaluating SCFGs for RNA compression and prediction and we explore best practices for improving the efficiency of the search for good grammars.

Moreover, we report results from an initial exploration of the space of grammars.
We find that the vast majority of grammars give rather poor compression,
but a very small number achieve substantial compression.
Among those, we could identify several new grammars that surpass the performance of similar-sized human expert grammars from the literature,
indicating that further improvements are likely to be possible and 
that the intuitive grasp of RNA structures even among domain experts has limitations.

The rest of this paper is structured as follows.
In \wref{sec:preliminaries}, we introduce basic notation and summarize how SCFGs can be used to represent RNA.
In \wref{sec:SRF}, we introduce our new normal form for SCFGs for RNA compression.
Our experimental setup and results are described in \wref{sec:results},
and we conclude in \wref{sec:conclusion}.
Datasets and code to produce figures and tables in this article are available online as supplementary
material: \url{https://www.wild-inter.net/publications/onokpasa-wild-wong-2024}; 
the code is available on GitHub: \url{https://github.com/evita35/better-grammars}.

\section{Preliminaries}
\label{sec:preliminaries}

We give a few basic definitions on strings and grammars,
before we introduce SCFGs as probabilistic models for RNAs.

We abbreviate $[a..b) = \{a,a+1,\ldots,b-1\}$.
For a string $S \in \Sigma^n$, we denote by $S[i]$ for $i\in[0..n)$ the $i$th character of $S$ (with 0-based indexing).  $S[i..j)$ denotes the substring $S[i]S[i+1]\ldots S[j-1]$.

\paragraph{RNA as strings}
An RNA sequence is a string of bases \texttt{A}, \texttt{C}, \texttt{G}, \texttt{U}.
Stable hydrogen bonds are possible between \texttt{A} and \texttt{U} resp.~\texttt{C} and \texttt{G}
(the Watson-Crick pairs) and to a lesser extent also between \texttt{G} and \texttt{U}.
(Pseudoknot-free) RNA (secondary) structures%
\footnote{%
	As is often done in the area, we do not consider structures with ``pseudoknots'' in this paper,
	\ie, we assume that all bonds are well nested.
}
can then be represented by the dot-bracket notation~\cite{Hofacker1994}:
a well-nested string over $\{\ssu,\sso,\ssc\}$ where a base pair is denoted by a matching pair of parentheses~``$\sso\,\ssc$'' and an unpaired base by ``$\ssu$'';
see \wref{fig:rna-example} for an example.
We use ``RNA'' as an abbreviation for ``a pair of an RNA sequence and its secondary structure''.
Formally, they are strings over pairs of characters (see also~\cite{OnokpasaWildWong2023}),
\eg, $\tao$ for base \texttt{A} in the RNA sequence
and \sso in the (dot-bracket representation of the) secondary structure.

\paragraph{Context-free Grammars}
Dot-bracket strings can be generated by a context-free grammar (CFG).
We use standard terminology for context-free grammars, see, \eg,~\cite{HopcroftMotwaniUllman2001}. 
All our derivations are leftmost derivations.
\ifarxiv{% arxiv version: give proper formal definitions
	\par
	A CFG is a tuple $(N,T,R,S)$ where
	$N$ and $T$ are finite sets of \emph{nonterminals} and \emph{terminals}, respectively,
	$R \subseteq N \times (N \cup T)^*$ is a finite set of \emph{production rules},
	and $S\in N$ is the \emph{start symbol}.
	A rule $(A,\gamma)\ \in R$ is written as $A \to \gamma$.
	We will use capital letters to denote nonterminals and lowercase letters for terminals.
	
	A \emph{leftmost application} of a rule $A\to\gamma$ in a sentential form $\alpha \in (N\cup T)^*$, provided $A$ is the leftmost nonterminal in $\alpha$, \ie, provided $\alpha = xA\beta$ for some $x\in T^*$ and $\beta \in (N\cup T)^*$, is the sentential form $\leftmost{A\to\gamma}{xA\beta} = x \gamma \beta$.
	A \emph{leftmost derivation} in $G$ is a sequence of rules $r_1,\ldots,r_t$ when the rules applied in sequence always lead to a well-defined leftmost derivation, \ie, with $\alpha_0 = S$ and $\alpha_i = \leftmost{r_i}{\alpha_{i-1}}$.
	We extend $\leftmost{}{\cdot}$ to sequences of rules, so write $\leftmost{r_1,\ldots,r_t}{\alpha_0} = \alpha_t$.
	All derivations in this work are leftmost derivations, we will therefore omit ``leftmost'' for brevity.
	We are mostly interested in \emph{terminal} (leftmost) derivations, \ie, derivations with $\alpha_t = w \in T^*$.
	The language of $G$ is the set of all $w\in T^*$ for which there is a (leftmost) derivation producing $w$.
}{}
We identify a derivation for word $w$ with the sequence of rules $r_1,\ldots,r_t$ used in the derivation;
we write $\leftmost{r_1,\ldots,r_t}{S} = w$ (for $S$ the start symbol) to indicate that rules $r_1,\ldots,r_t$, successively applied starting with $S$, produce $w$.

\subsection{Stochastic Context-free Grammars}

A \emph{stochastic context-free grammar} (SCFG) is a tuple $G=(N,T,R,S,P)$
such that
$(N,T,R,S)$ is a CFG and 
for every $A \in N$, 
$P: R \to[0,1]$ induces a probability distribution over the set of rules with left-hand side $A$.

\paragraph{SCFG as probabilistic models}
The probability of a derivation $r_1,\ldots,r_t$ (a sequence of rules from $R$) in the grammar $G$ is the product of the probabilities of all used rules: $\Prob{r_1,\ldots,r_t} = \prod_{i=1}^t P(r_i)$.
This corresponds to the probability of obtaining this derivation in the random process, where starting with $S$, in each time step,
we choose a random replacement for the leftmost nonterminal $A$ in the current sentential form. For that, we sample one of the rules $A\to\gamma$ with probabilities according to $P$ and, conditionally on having left-hand side $A$, independent of the past choices.

We define the probability $\Prob{w}$ of a \emph{word} $w$ as 
the sum of the probabilities of its derivations.
\ifarxiv{
    \[
    		\Prob{w} 
    	\wwrel= 
    		\sum_{\substack{
    			r_1,\ldots,r_t:\;
    			\leftmost{r_1,\ldots,r_t}{S} = w}} 
    			\Prob{r_1,\ldots,r_t} 
    	\;.
    \]
    The sum is understood to range of all leftmost derivations (of arbitrary length).
}{}%
We also define the \emph{Viterbi value} $V(w)$ of $w$, the probability of the most likely derivation of $w$:
\[
		V(w)
	\wwrel= 
		\max_{
                %\substack
                {
			     r_1,\ldots,r_t:\;
                %\\
			     \leftmost{r_1,\ldots,r_t}{S} = w}
                } 
			\Prob{r_1,\ldots,r_t} 
	\;.
\]
If $G$ is unambiguous, there is only one derivation and we have $\Prob{w} = V(w)$.

\paragraph{Derivations as representations}

If $G$ is known (by convention or because it has been stored explicitly), a leftmost derivation $d = (r_1,\ldots,r_t)$ of a word $w = \leftmost{d}{S}$, is an encoding for $w$: the original word can always be reconstructed from $d$ by (leftmost) application of the rules starting with $S$.  
This is the basis for our RNA encoding~\cite{OnokpasaWildWong2023}.
Note that the grammar is not required to be unambiguous for that, although it seems plausible that unambiguous grammars would yield more effective compression.

\paragraph{Probabilistic Parsing}

Given a SCFG $G = (N,T,R,S)$ and a word $w\in T^*$ in the language of $G$, a probabilistic parser determines
a \emph{Viterbi derivation}, \ie, a most likely derivation for $w$: $\arg\max \{ \Prob{d} : \leftmost{d}{S} = w\}$.

The theory of such parsers is well established and does not require $G$ to have any specific normal form~\cite{Goodman1999,Goodman1998}. However the resulting general algorithms are rather intricate;
for example, chain rules can require to 
(symbolically) solve infinite summations for correct stochastic parsing~\cite{Goodman1998}. 
But such grammars immediately allow an infinite number of leftmost derivations for one word;
since our compression methods specify a single  derivation, such ambiguity is counterproductive for compression.

A normal form such as \emph{Chomsky normal form (CNF)} (all rules of type $A\to c$ or $A\to BC$)
can simplify parsing dramatically; 
here even a stochastic parser remains a simple (bottom-up) dynamic-programming algorithm (stochastic CYK)~\cite{Goodman1999}.

Unfortunately, CNF is inconvenient for expressing the complementarity of paired bases in an RNA.
We therefore start by proposing a new normal form for our grammars in \wref{sec:SRF}.

\subsection{SCFG-based Joint RNA Compression}

Our formalism from~\cite{OnokpasaWildWong2023} unifies grammars for encoding an RNA structure, joint RNA (sequence and structure), and predicting the structure from the sequence.
We specify a grammar by giving the rules, \eg,
$
	S \to \sso S \ssc \mid \ssu \mid S S
$;
here `$\mid$' separates the right-hand sides of rules with the same nonterminal on the left.
This represents 3 different types of grammars: (1) As is, it is a grammar for deriving/representing just the RNA structure (in dot-bracket notation). We call these the \emph{secondary-structure grammar}. (2) We can expand the secondary-structure grammar to an \emph{RNA grammar} by replacing all rules with $\ssu$ by 4 rules, where $\ssu$ is replaced by $\tau$, $\tcu$, $\tgu$, and $\tuu$, respectively, and all rules with a pair of $\sso$ and $\ssc$
by 6 rules, where \sso{}/\,\ssc is replaced by
$\tao$/$\tuc$,
$\tco$/$\tgc$,
$\tgo$/$\tcc$,
$\tgo$/$\tuc$,
$\tuo$/$\tac$, and 
$\tuo$/$\tgc$, respectively.
(Or, for handling datasets with non-canonical base pairs, even generate all $4\times 4$ combinations of bases instead of just those 6).
For brevity, we will use the short notation for the secondary-structure grammar, but we actually work with the expanded RNA grammar for compression.

(3) The third type, the \emph{prediction grammar}, has the same structure as the RNA grammar, but when using it in a parser, we only look at the first entry of each pair. 
Thereby we can use it to compute a (most likely) derivation for a given RNA sequence; the pairs matched with the bases in the sequence automatically give us the RNA structure corresponding to this derivation. 
We can thus use a Viterbi parse to obtain the most likely structure for a given sequence using the prediction grammar.

\paragraph{Rule-probability models}
When using an RNA grammar $G$ for compressing an RNA sequence and structure pair $w$, we first determine 
a derivation for $w$ in $G$. For the encoding of $w$ given grammar $G$, we now need to specify probabilities for the rules.
As in~\cite{OnokpasaWildWong2023}, we focus on two options here: (a) a \emph{static} rule-probability model, where
we determine probabilities from counting how often each rule is used on a training dataset, and
(b) an \emph{adaptive} rule-probability model, where we keep running counts of rule occurrences (starting at 1, \ie, a uniform prior) in the already encoded prefix of the derivation.
For the actual binary encoding, we employ arithmetic coding~\cite{Witten1987} to store the next rule. Note that the left-hand side is always known; initially it is the start symbol  and then, inductively, the leftmost nonterminal in the current sentential form.
More details and a worked example are given in~\cite[\S3]{OnokpasaWildWong2023}.

\section{Stochastic RNA Normal Form for Grammars}
\label{sec:SRF}

We consider SCFGs in a specific normal form, the \emph{Stochastic RNA Form (SRF)}.
It takes inspiration from existing expert SCFG designs used for RNA structure prediction~\cite{DowellEddy2004,LiuYangChenBuZhangYe2008,NebelScheid2011}.
Our normal form assumes a total order on the nonterminals.  
To simplify notation in this section, assume without loss of generality that $N = \{A_1,\ldots,A_k\}$; 
we define $A_i < A_j$ if and only if $i<j$.

A SCFG $G=(N,T,R,A_k,P)$ is in \emph{Stochastic RNA Form} if each of its rules has one of the following forms:

\medskip

\noindent
\plaincenter{\fbox{
    (i)\; $A_i \to A_j A_l$
\quad\;
    (ii)\; $A_i \to \ssu$
\quad\;
    (iii)\; $A_i \to \sso A_j \ssc$
\quad\;
    (iv)\; $A_i \to A_j$ and $j<i$ 
}}

\medskip

\noindent
The ordering constraint on type (iv) rules ensures that there is a finite number of derivations for every word
and that we can retain a total order on subproblems in parsing (see below).
Apart from making parsing more efficient, 
the Stochastic RNA Form makes it trivial to ensure that grammars produce valid RNA structures.
It therefore massively reduces the search space in our exhaustive search 
by excluding many invalid grammars.

The Stochastic RNA Form is chosen to be as expressive as possible, fixing only features that all stable RNA structures share.
One tacit assumption we impose on RNA structures is that they have no empty hairpin loops, \ie, no subword ``\sso\,\ssc''.  
Such a bond is indeed impossible due to physical limitations; it does get reported in databases, but rarely so (and likely erroneously).

Given a grammar in SRF, we can adapt the probabilistic CYK parser~\cite{Goodman1999} to our grammars as follows:
For that, we denote by $V_A(w[i..j))$, for $A\in N$ and $0\le i < j \le n=|w|$ the probability of the most likely derivation of $w[i..j)$ when starting with $A$.
We then have $V(w) = V_S(w[0..n))$ for $S$ the start symbol of $G$ and obtain the recursive equations following the allowed rule types:
\begin{gather*}
		V_A(w[i..i+1) )
	\wrel=
		\begin{dcases*}
		P(A\to w[i]) & if $A\to w[i] \rel\in R$ (ii)\\
		0 & otherwise
		\end{dcases*}
\\[1ex]
		V_A(w[i..j) )
	\wrel=
		\max\begin{dcases*}
			\max_{\substack{k \in [i+1..j)\\A\to BC \rel\in R}} 
					P(A\to BC) \, V_B(w[i..k)) \, V_C(w[k..j))  & (i) \\
			\max_{A\to w[i] B w[j-1] \rel\in R} 
					P(A\to w[i] \, B \, w[j-1])\,  V_B(w[i+1..j-1))  & (iii) \\
			\max_{A\to B \rel\in R} 
					P(A\to B) \, V_B(w[i..j))  & (iv) \\
		\end{dcases*}
\end{gather*}

The ordering constraint on type (iv) rules implies that the subproblems $V_A(w[i..j))$ can be totally ordered by $(\ell, A)$ for $\ell=j-i$ the length of the produced subword and $A$ the nonterminal, allowing for an efficient bottom-up dynamic-programming parser.

\section{Methodology and Results}
\label{sec:results}

\subsection{Exhaustive exploration}

We first report on the results of an exhaustive exploration of all small stochastic context-free grammars in Stochastic RNA Form.
The goal is to shed light on the following questions: 
\textit{Does RNA favor a specific shape of grammars? If so, which?
How different is the compression ability of different grammars of the same size?
Are the human-expert chosen grammars best possible for their size?}

For this, we implemented an exhaustive generation algorithm that can iterate over all possible grammars in Stochastic RNA Form by constructing for a given number of nonterminals all possible rules in SRF.
Then we can iterate over all possible subsets, or all subsets of a given size to generate all possible SRF grammar with these parameters.
Note that for $k$ nonterminals, the total number of possible SRF rules is 
$k^3$ type (i) rules, $k$ type (ii) rules, $k^2$ type (iii) rules and $\binom k2$ type (iv) rules.

\begin{table}[tb]
	\centering\footnotesize
	\adjustbox{max width=\linewidth}{
	\begin{tabular}{rrrrrr}
	\toprule
		   \textbf{\#NTs} & \textbf{\#rules} & \textbf{\# SRF grammars} & \multicolumn{2}{r}{\textbf{\#parsing gr.\; (\%)}} & \textbf{best bits-per-base} \\
	% NOTE: #parsing is #grammars reaching stage 3, except for 3NTs >=7 rules, where it is passed stage 1
\midrule
		                1 &                3 &                        1 &           1 &                               100\% &                      3.6241 \\
	\midrule
		                2 &                1 &                       15 &           0 &                                 0\% &                         --- \\
		                2 &                2 &                      105 &           0 &                                 0\% &                         --- \\
		                2 &                3 &                      455 &           1 &                               0.2\% &                      3.6890 \\
		                2 &                4 &                     1365 &          28 &                                 2\% &                      3.1424 \\
		                2 &                5 &                     3003 &         201 &                                 7\% &                      2.9969 \\
		                2 &                6 &                     5005 &         783 &                                16\% &                      2.9762 \\
		                2 &                7 &                     6435 &        1831 &                                28\% &                      2.9927 \\
		                2 &                8 &                     6435 &        2801 &                                44\% &                      3.0088 \\
		                2 &                9 &                     5005 &        2953 &                                59\% &                      3.0516 \\
		                2 &               10 &                     3003 &        2198 &                                73\% &                      3.0668 \\
		                2 &               11 &                     1365 &        1158 &                                85\% &                      3.0856 \\
		                2 &               12 &                      455 &         424 &                                93\% &                      3.1229 \\
		                2 &               13 &                      105 &         103 &                                98\% &                      3.3456 \\
		                2 &               14 &                       15 &          15 &                               100\% &                      3.5364 \\
		                2 &               15 &                        1 &           1 &                               100\% &                      3.8076 \\
	%	\midrule
		%
	\midrule
		                3 &                1 &                       42 &           0 &                                 0\% &                         --- \\
		                3 &                2 &                      861 &           0 &                                 0\% &                         --- \\
		                3 &                3 &                  11\,480 &           1 &                           0.00001\% &                      3.6891 \\
		                3 &                4 &                 111\,930 &          71 &                           0.00001\% &                      3.1424 \\
		                3 &                5 &                 850\,668 &        2015 &                             0.002\% &                      2.9866 \\
		                3 &                6 &              5\,245\,786 &     33\,170 &                             0.006\% &                      2.6620 \\
		                3 &                7 &             26\,978\,328 &    377\,522 &                             0.013\% &                      2.5495 \\
		                3 &                8 &            118\,030\,185 & 3\,212\,691 &                             0.027\% &                      2.5582 \\
	\bottomrule
	\end{tabular}
	}
	\caption{
		Overview of the overall number of grammars of certain sizes and the number of grammars that are able to parse all RNA in our benchmark dataset.
		The best bits-per-base gives an indicative compression performance (adaptive rule-prob model on the 10\% sample of ``benchmark'').
		Note that the number of possible rules for 2 nonterminals is 15 and for 3 nonterminals is 42.
	}
	\label{tab:parsing-grammars}
\end{table}

The number of grammars is large ($> 2^{k^3}$) even for moderate $k$; however many grammars do not even allow to parse all RNA structures (see \wref{tab:parsing-grammars}) and can be discarded quickly.
For that, we use a tiny dataset ``parsable'' of short RNA structures; any grammar that fails to parse this dataset is skipped.
For the remaining grammars, we determine the normalized compressed size, \ie, the number of bits in the compressed representation divided by the number of bases of the RNA;
both using adaptive and static rule-probability models.
The dataset we use is the ``benchmark'' dataset from Dowell and Eddy~\cite{DowellEddy2004}. As an efficient first filter we reduce it to a randomly chosen 10\% subsample; we determined in preliminary experiments that this predicts the bits-per-base value on the entire dataset to within $\pm1\%$.  
We hence determine the best grammars from this 10\% subsample and then evaluate the most successful grammars on the full benchmark dataset.

\subsection{Distribution of compression ability}

\begin{figure}
	\centering
	\ifarxiv{%
		\includegraphics[width=1\linewidth]{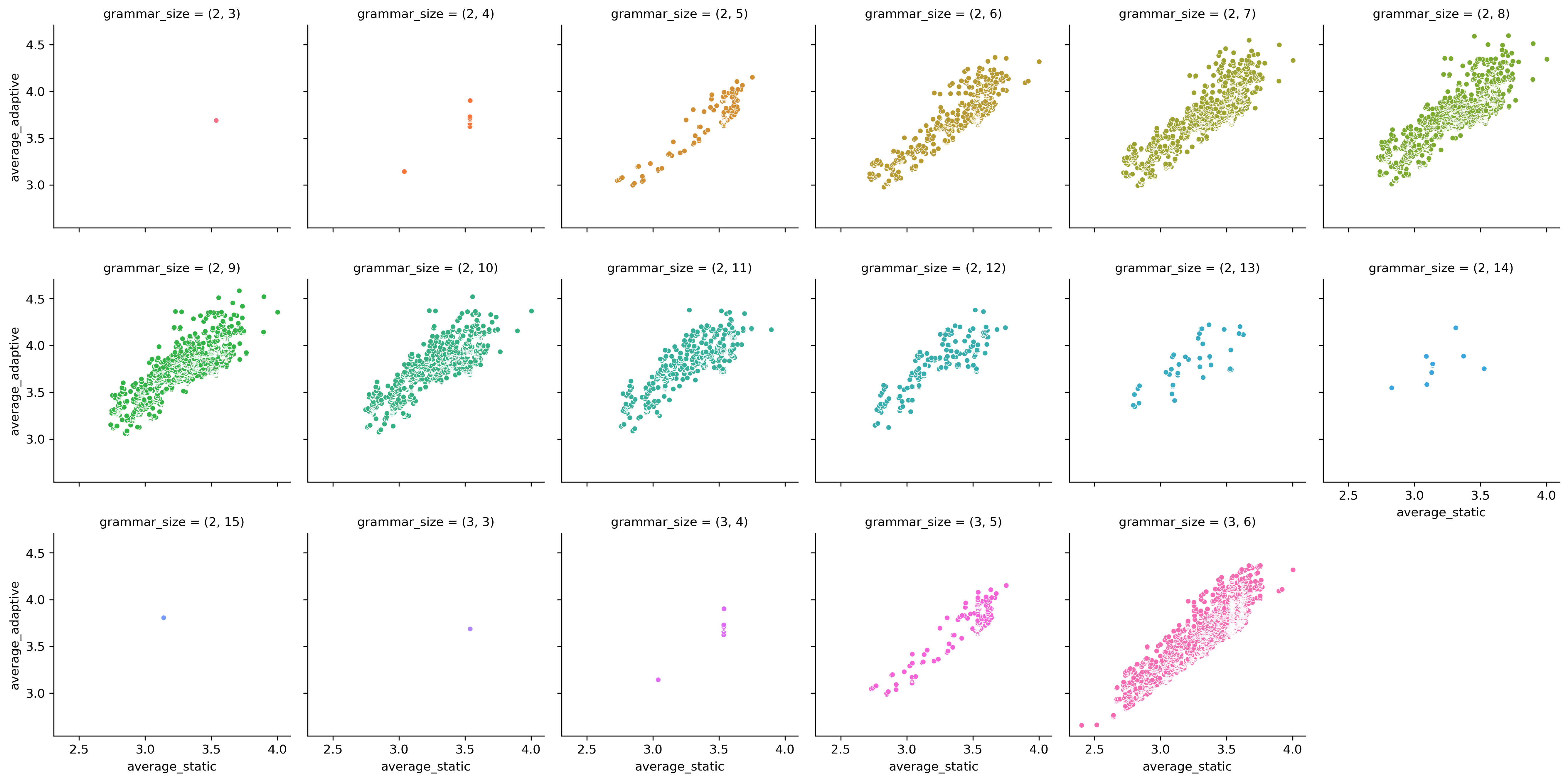}
	}{%
		\includegraphics[width=.9\linewidth]{pics/bits-per-base-exhaustive-static-vs-adaptive}
	}%
		\caption{%
			Normalized average compressed size (in bits per base) for all grammars of the given size (\#NTs (Nonterminals), \#rules) on 10\% sample of the ``benchmark'' dataset from~\cite{DowellEddy2004}. Each dot is one grammar; the $x$-coordinate is using the static rule-probability model, with rule counts on the same dataset; the $y$-coordinate uses the adaptive rule-probability model.
		}
	\label{fig:exhaustive-by-size}
\end{figure}

\begin{figure}
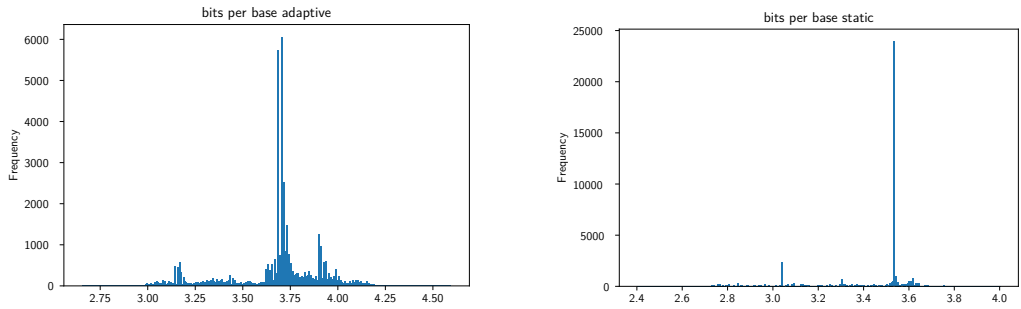

	\plaincenter{
		\resizebox{.4\linewidth}!{\input{pics/bits-per-base-exhaustive-adaptive-histogram.pgf}}\hfill
		\resizebox{.4\linewidth}!{\input{pics/bits-per-base-exhaustive-static-histogram.pgf}}
	}
	\caption{%
		Histogram of the normalized average compressed size (in bits per base) for all grammars from \wref{fig:exhaustive-by-size} on the 10\% subsample of the benchmark dataset from Dowell and Eddy~\cite{DowellEddy2004} using the adaptive (left) resp.\ static (right) rule-probability model. 
	}
	\label{fig:exhaustive}
\end{figure}

\begin{figure}
	\centering
	\scalebox{.9}{
	\begin{minipage}[t]{\textwidth}
		\footnotesize\let\oldsubsection\subsection
		\def\section#1{}
		\def\subsection*#1{\columnbreak#1\vspace{1ex}}
			\begin{multicols}{6}
				\input{optimal-grammars-appendix-grammars}
			\end{multicols}
	\end{minipage}}
	\caption{%
		Newly identified grammars; $G^*_{k,r}$ is the best grammar with $k$ NTs (Nonterminals) and $r$ rules from exhaustive search; \smash{$G^\dag_{k,r}$} is the best grammar we found with random search.
	}
	\label{fig:our-new-grammars}
\end{figure}

Using each possible small grammar (all grammars with 2 nonterminals, and all grammars with 3 nonterminals and at most 6 rules) to compress the 10\% sample of the benchmark dataset, we obtain the normalized compressed size (in bits per base) using the adaptive and static rule-probability models.  
\wref{fig:exhaustive-by-size} shows scatter plots of these results, split by grammar size. 
There is a clear correlation of the two measures, meaning that grammars mostly live on a single scale from better to worse compression approximately reflecting both models.

Moreover, the vast majority of grammars give rather poor compression.
This is even more visible in \wref{fig:exhaustive} which shows the distribution of bits-per-base for all grammars up to 3 nonterminals and 6 rules (same data as in \wref{fig:exhaustive-by-size}).
We expect the normalized compressed size to be at least 2 bit per base, since the primary structure of RNA is not known to be substantially compressible (and thus needs roughly 2 bits per character), and we need some additional information to encode the structure on top of that.
The vast majority of grammars just realize a compressed size around $\lg(4\cdot 3)\approx 3.58$, which corresponds to storing each of the pairs of terminals (4 bases, 3 structure symbols) independently with uniform likelihood.
Note that this is (approximately) the compression achieved by the trivial grammar with a single nonterminal and the three rules $A\to \sso A \ssc \mid \ssu \mid A A$.
It seems hence indeed the case that most grammars are not able to pick up the structure of RNA at all, and only a very small number of grammars achieve substantially better compression than almost all other grammars.
\wref{fig:our-new-grammars} shows some of these.

\subsection{Random search}

Despite an increasing fraction of grammars not parsing all RNA and most grammars only giving trivial compression,
compression quality does increase with (moderate) increase in grammar size (see \wref{tab:parsing-grammars} and \wref{fig:exhaustive-by-size}). It is therefore desirable to be able to search among larger grammars than accessible via exhaustive exploration.
As a simple first step towards that, we implemented a random grammar explorer that repeatedly generates random grammars (from the grammars of a given size) and keeps track of the top $m$ grammars ever encountered.
Again, to obtain any meaningful efficiency, we first check grammars for the ability to parse a tiny dataset of artificial RNA.
Among those who parse that correctly, we evaluation their compression ability on a small dataset of short RNA and those who perform on this at least as good as the worst of the current top $m$ grammars, we evaluate on the benchmark dataset.  The top $m$ grammars are always chosen based on the benchmark dataset.
(We confirmed on the small grammars where we have exhausted all grammars that this process indeed identified the overall best performing grammars.)
We ran this random exploration for grammars with 3--10 nonterminals and different numbers of rules, together exploring several billions of grammars.
The best grammar found in that process was $G^\dag_{6,10}$, see \wref{fig:our-new-grammars} (right). Note that $G^\dag_{6,10}$ contains 2 nonterminals that are dead ends for any derivation: $A_3$ has no rules, and hence cannot ever be replaced. $A_2$ only has rules involving $A_3$ and hence can likewise never be resolved to terminals.
Removing those rules (and nonterminals) gives $G^\dag_{6,10}{}'$.

\subsection{Comparison with expert-curated grammars}

We compare the results from newly found grammars with the expert-curated RNA grammars collected in the literature in \wref{tab:vs-de04}; the grammars can be found in \wref{fig:our-new-grammars} resp.\ the appendix of~\cite{OnokpasaWildWong2023}.
Although by a narrow margin, the best grammar for adaptive rule probabilities known is our new grammar $G^\dag_{6,10}{}'$, clearly demonstrating that human-expert grammars are not necessarily best possible!

\begin{table}
	\centering
	\adjustbox{max width=\linewidth}{\footnotesize\begin{tabular}{lrrrrr}
	\toprule
		\textbf{Grammar}                        & \textbf{adaptive} & \textbf{static} & \textbf{\#NTs} & \textbf{\#rules} & \textbf{grammar size} \\
	\midrule
		grammar-1NT                             &            3.6241 &          3.4731 &              1 &                3 &                     3 \\
	\midrule
		$G_{L'}$ (Liu et al.)                   &            3.1229 &          3.0201 &              2 &                4 &                    18 \\
		$G^*_{2,5}$ (new)                       &            2.9699 &          2.8200 &              2 &                5 &                    19 \\
		$G^*_{2,6}$ (new)                       &            2.9494 &          2.8011 &              2 &                6 &                    20 \\
	\midrule
		$G_5$ (Dowell, Eddy)                    &            2.8368 &          2.7423 &              3 &                6 &                    32 \\
		$G_3$ (Dowell, Eddy)                    &            2.5804 &          2.4549 &              5 &               11 &                    69 \\
		$G_6$ (Dowell, Eddy)                    &            2.4957 &          2.3687 &              5 &                9 &                    61 \\
		$G_4$ (Dowell, Eddy)                    &            2.7138 &          2.5974 &              6 &               11 &                    78 \\
		$G_1$ (Dowell, Eddy)                    &            3.0779 &          2.8956 &              6 &               13 &                    87 \\
		$G_2$ (Dowell, Eddy)                    &            2.9723 &          2.5525 &             18 &              296 &                  1742 \\
		$G_7$ (Dowell, Eddy)                    &            2.6343 &          2.3333 &             38 &              321 &                  2883 \\
		$G_8$ (Dowell, Eddy)                    &            2.7213 &          2.4561 &             39 &              322 &                  2926 \\
	\midrule
		$G_S$ (Nebel, Scheid)                   &            2.5045 &          2.2876 &            108 &              244 &                  3396 \\
	\midrule
		$G^*_{3,6}$ (new)                       &            2.6329 &          2.3797 &              3 &                6 &                    32 \\
		$G^*_{3,7}$ (new)                       &            2.5287 &          2.3878 &              3 &                7 &                    34 \\
		$G^\dag_{6,10}$ (new)                   &            2.5091 &          2.3835 &              6 &               10 &                    73 \\
		$G^\dag_{6,10}{}'$ (new)                  &            2.4902 &          2.3835 &              4 &               7 &                    45 \\
	\bottomrule
	\end{tabular}}
	\caption{%
		Normalized compressed size (bits per base) for some newly found small grammars
		and the grammars from~\cite{DowellEddy2004,LiuYangChenBuZhangYe2008,NebelScheid2011}.
		All results are for the ``benchmark'' dataset from~\cite{DowellEddy2004}.
		``grammar size'' is the \#bits needed to encode the grammar using the following scheme: encode $k$ (\#NTs) in Elias gamma code, then $r$ (\# rules) in binary and finally the size-$r$ subset of rules using an enumeration of all size-$r$ subsets.
	}
	\label{tab:vs-de04}
\end{table}

Two expert grammars are in the range where exhaustive exploration is still feasible and both are \emph{not} the best possible grammar, even in their (flyweight) category: 
$G_{L'}$ by Liu et al.~\cite{LiuYangChenBuZhangYe2008} is surpassed by a fair margin (2.95 vs 3.12 bits per base) by other grammars with 2 nonterminals. It is best possible, though if we insist on exactly 4 rules.
For $G_5$ from Dowell and Eddy~\cite{DowellEddy2004}, we find a better grammar of the same size, again with a substantial improvement of compression ability.

\section{Conclusion}
\label{sec:conclusion}

We formulated the competition for the best joint RNA compression grammar and gave first improvements on the best known grammars.
Unfortunately, the combinatorial explosion of possible grammars and the fact that most grammars do not compress RNA meaningfully
turns the problem into the search for a needle in a haystack. 
Further progress in future work will likely have to come from heuristics, potentially including learning systems. 

The fact that the expert grammars are (in several ways) not best possible indicates that there are still unexplored patterns in RNA structures to be understood. 
The contest to find the best grammar for adaptive rule probabilities in particular has the potential to deepen our understanding of RNA structures and might lead to interesting compression methods on the path towards optimal grammars for RNA compression and prediction.

	\myacknowledgements
%

%\clearpage
% !BIB program = bibtex
\ifdcc{\section*{References}}{}
\begin{small}
\bibliography{references}
\end{small}

\ifsubmission{}{
\ifproceedings{}{
	
	\clearpage
	\appendix
	\ifkoma{\addpart{Appendix}}{}
	
	\ifsubmission{}{\section{New Grammars}}
\label{app:grammars}

\subsection*{$G^*_{2,6}$}

      $A_0 \rightarrow \ssu$ 
    \\$A_0 \rightarrow A_1 A_0$ 
    \\$A_0 \rightarrow A_1 A_1$ 
    \\$A_1 \rightarrow A_0 A_1$ 
    \\$A_1 \rightarrow \sso A_1 \ssc$ 
    \\$A_1 \rightarrow A_0$ 

\subsection*{$G^*_{2,5}$ }

      $A_0 \rightarrow \ssu$ 
    \\$A_1 \rightarrow A_1 A_1$ 
    \\$A_1 \rightarrow A_0 A_1$ 
    \\$A_1 \rightarrow \sso A1 \ssc$ 
    \\$A_1 \rightarrow \ssu$

\subsection*{$G^*_{3,6}$}

  $A_0 \rightarrow \ssu $
\\$A_0 \rightarrow \sso A_2 \ssc $
\\$A_1 \rightarrow A_0 $
\\$A_2 \rightarrow \sso A_2 \ssc $
\\$A_2 \rightarrow A_1 $

\subsection*{$G^*_{3,7}$}

  $A_0 \rightarrow \ssu$
\\$A_0 \rightarrow \sso A_0 \ssc$ 
\\$A_0 \rightarrow A_1 A_2 $
\\$A_1 \rightarrow \ssu $
\\$A_1 \rightarrow A_0 $
\\$A_2 \rightarrow A_1 $
\\$A_2 \rightarrow A_1 A_2 $

\subsection*{$G^\dag_{6,10}$}

  $A_0 \rightarrow \sso A_5 \ssc $
\\$A_1 \rightarrow \ssu $
\\$A_1 \rightarrow A_0 $
\\$\color{black!50}A_2 \rightarrow A_3 A_5 $
\\$\color{black!50}A_2 \rightarrow A_4 A_3 $
\\$A_4 \rightarrow A_1 $
\\$A_4 \rightarrow A_4 A_1 $
\\$A_5 \rightarrow A_0 $
\\$\color{black!50}A_5 \rightarrow A_0 A_2 $
\\$A_5 \rightarrow A_4 $

}}

\ifdraft{
	\clearpage
	\part*{Notes-to-self}
	\printnotestoself
}{}

\end{document}